\documentclass[aps,prl,twocolumn,floatfix,10pt,groupedaddress]{revtex4-2}

\usepackage{graphicx,amsmath,amssymb,ulem}
\usepackage{textcomp}
\usepackage{xcolor}
\usepackage{hyperref}
\usepackage{color}
\usepackage{epstopdf}
\usepackage{nicefrac}
\usepackage{gensymb}
\usepackage{placeins}
\usepackage{xspace}
\usepackage{lineno}

\def\40K{$^{40}$K}
\def\K{$^{39}$K}
\def\Na{$^{23}$Na}
\def\Rb{$^{87}$Rb}
\def\Cs{$^{133}$Cs}
\def\NaK{\Na\K}

\def\ket#1{\mathinner{|{#1}\rangle}}


\begin{document}

\title{Probing  photoinduced two-body loss  of ultracold non-reactive bosonic \Na\Rb\, and \Na\K\, molecules}

\author{Philipp~Gersema$^1$}
\author{Kai K.~Voges$^1$}
\author{Mara Meyer zum Alten Borgloh$^1$}
\author{Leon Koch$^1$}
\author{Torsten~Hartmann$^1$}
\author{Alessandro~Zenesini$^{1,2}$}
\author{Silke~Ospelkaus$^1$}
\email{silke.ospelkaus@iqo.uni-hannover.de}
\affiliation{$^1$Institut f\"ur Quantenoptik, Leibniz Universit\"at Hannover, 30167~Hannover, Germany\\
$^2$INO-CNR BEC Center and Dipartimento di Fisica, Universit\`{a} di Trento, 38123 Povo, Italy
}

\author{Junyu Lin$^3$}
\author{Junyu He$^3$}
\author{Dajun Wang$^3$}
\email{djwang@cuhk.edu.hk}
\affiliation{$^3$Department of Physics, The Chinese University of Hong Kong, Hong Kong, China}

\date{\today}

\begin{abstract}
	We  probe photo-induced loss for chemically stable bosonic \Na\Rb\, and \Na\K\ molecules in chopped optical dipole traps where the molecules spend a significant time in the dark. We expect the effective two-body decay to be largely suppressed in chopped traps due to the small expected complex lifetimes of about 13\,$\mu$s and 6\,$\mu$s for \Na\Rb\, and \Na\K\, respectively. However, instead we do observe near-universal loss even at the lowest chopping frequencies we can probe. Our data thus either suggest  a drastic underestimation of the complex lifetime by at least one to two orders of magnitude or a so far unknown loss mechanism.

\end{abstract}

\maketitle

The field of ultracold heteronuclear molecules has gained a lot of attention in recent years~\cite{Bohn1002}. Nowadays, full simultaneous control over the molecules' external and complex internal structure including  electronic vibrational, rotational and hyperfine degrees of freedom is within reach. This is opening many new research opportunities such as quantum simulation of new many-body states with tunable anisotropic long-range interactions \cite{PolarMoleMoses,KRbDegenerate, KRbSpin},  the possibility for quantum-state controlled chemistry \cite{Ospelkaus2010,WangCollisions}, quantum computation \cite{DeMilleQuantComp}, and precision measurements \cite{269,Nature562ImprovedElectron}.

Currently, the most advanced experimental molecular system is given by ultracold gases of bialkali rovibrational ground state polar molecules. The first creation of an ultracold ensemble of polar ground-state molecules with high phase-space density has been achieved  with the fermionic $^{40}$K$^{87}$Rb molecules at JILA~\cite{Ni231, Ospelkaus2010}. Ultracold clouds of these molecules suffered from strong two-body losses at the universal limit which were later explained by an exothermic chemical reaction taking place during the collision between two molecules, $\text{2\,KRb}\rightarrow \text{K}_\text{2}+ \text{Rb}_\text{2}$\cite{Ospelkaus2010, Inelastic2010, PhysRevA.82.042707, Hu1111}.

 Since then considerable effort has been devoted to the production of chemically stable ground-state molecules. Molecules such as $^{87}\text{Rb}^{133}\text{Cs}$  \cite{GsDiMo87Rb133Cs2014Grimm,GsDiMo87Rb133Cs2014Cornish},$^{23}\text{Na}^{87}\text{Rb}$  \cite{NaRb1} and fermionic $^{23}\text{Na}^{40}\text{K}$  \cite{GsDiMo23Na40K2015, GsDiMo23Na40KFerm2018Bloch} as well as bosonic $^{23}\text{Na}^{39}\text{K}$ \cite{PhysRevLett.125.083401} have been successfully prepared in the laboratory.
However, independently of their fermionic or bosonic nature, mass or different experimental circumstances, in all of these experiments molecular two-body losses close to the universal scattering limit have been observed \cite{GsDiMo87Rb133Cs2014Grimm,NaRb1,GsDiMo23Na40K2015,PhysRevLett.125.083401}. 
In 2013 Mayle \textit{et al.} \cite{PhysRevA.87.012709} introduced the concept of long-lived four-body collisional complexes. During the collision of two non-chemically reactive ground-state molecules, the high density of resonant states leads to an enhanced scattering time, in which the two molecules stick together (sticky collisions) \cite{PhysRevA.85.062712,PhysRevA.87.012709}. Eventually these complexes are expected to break up and the  complexes convert back to molecules  in their initial quantum states. Unless the complexes get lost during their sticking time, this process should not result in molecular loss.  However, different loss mechanisms of the complexes have been discussed in the literature. Mayle \textit{et al.} calculated the complex lifetime to be long enough for a collision of the complex with another molecule during the sticking time. This collision would result in an effective loss of three molecules \cite{PhysRevA.87.012709}. 
In 2019 Christianen \textit{et al.} estimated a 2 to 3 orders of magnitudes lower density of states and associated lifetime of the complexes, making the collisional loss of four-body complexes far less likely \cite{PhysRevLett.123.123402}.
Instead Christianen \textit{et al.}  proposed photoinduced loss of the complexes to excited tetramer states by light of the optical dipole trap (ODT) as the dominant loss mechanism \cite{PhysRevLett.123.123402}. The complex excitation rates have been estimated to be fast enough to induce a loss rate of molecules identical to the complex formation rate. 

Recently, this hypothesis has been supported by experiments with bosonic \Rb\Cs\ molecules confined in a modulated (chopped) ODT \cite{PhysRevLett.124.163402}. As the ODT light strongly saturates the optical transitions of the four-body complexes \cite{PhysRevLett.123.123402}, photoinduced loss has been observed to be partially suppressed when applying square-wave modulations to the trap intensity, such that the molecules spend parts of the modulation cycle in the dark. Furthermore, by varying the frequency of the modulation, the  complex lifetime has been measured to $\tau_{c,\text{RbCs}}=0.53(6)\text{\,ms}$ \cite{PhysRevLett.124.163402}, which is within a factor 2 of the value $\tau_{c,\text{RbCs}}=0.253\text{\,ms}$ predicted by Christianen \textit{et al.} \cite{PhysRevA.100.032708}. Additionally, the complex lifetime estimate has been confirmed in a measurement of the transition intermediate complex lifetime with chemically reactive \40K\Rb\ molecules~\cite{liu_photo-excitation_2020}. It is of interest to expand the investigation of chopped ODT measurements to different non-chemically reactive molecular species. This allows to validate that the light excitation of complexes in the ODT is indeed the dominating loss mechanism, to measure the complex lifetimes and therefore the density of resonant states and to compare these to the theoretical predictions.

In this paper, we probe photoinduced collisional loss of ultracold chemically stable bosonic \Na\Rb\, and \Na\K\, ground-state molecules using chopped ODTs.  Given the short predicted complex lifetime of $13\,\mu$s and $6\,\mu$s for \Na\Rb\, and \Na\K\, respectively~\cite{PhysRevA.100.032708}, it is easy to enter a regime where the dark time $t_\textrm{d}$ of the chopped traps is more than ten times longer than the expected complex lifetime $\tau_\textrm{c}$. Assuming that photoinduced two-body loss is indeed the dominating decay mechanism, this should result in a largely suppressed two-body decay in chopped traps.  

Surprisingly and in stark contrast to our expectations, we do not find such a strong suppression. Our data is instead consistent with near-universal loss even at the lowest modulation frequencies $f_\textrm{mod}$ and largest $t_\textrm{d}$ we can probe and thus inconsistent with the short predicted complex lifetimes.  This suggests that the recent picture of dominant loss through complex formation and photoinduced loss and/or the predicted complex lifetime is still incomplete.

Assuming the dominant loss mechanism for ultracold non-chemically reactive molecules to be sticky collisions followed by photoinduced complex excitation, we can model the loss dynamics of molecules in continuous-wave (cw) and chopped ODTs \cite{PhysRevLett.124.163402}: 
\begin{equation}
\begin{split}
\dot{n}_m&=-k_2n_m^{2}+\dfrac{2}{\tau_c}n_c,\\
\dot{n}_c&=+\dfrac{1}{2}k_2n_m^{2}-\dfrac{1}{\tau_c}n_c-k_lI(t)n_c,
\end{split}
\label{eq}
\end{equation}
where $n_m$ is the molecule density, $n_c$ is the complex density, $k_2$ is the two-body loss rate coefficient measured in a continuously operated ODT , $k_l$ the excitation rate of the complexes per unit intensity $I(t)$ of the applied laser field.
 We assume that $k_lI(t)>>1/\tau_c$ in the ODT \cite{PhysRevLett.123.123402}.
\begin{figure}[t]
\includegraphics[width=1\columnwidth]{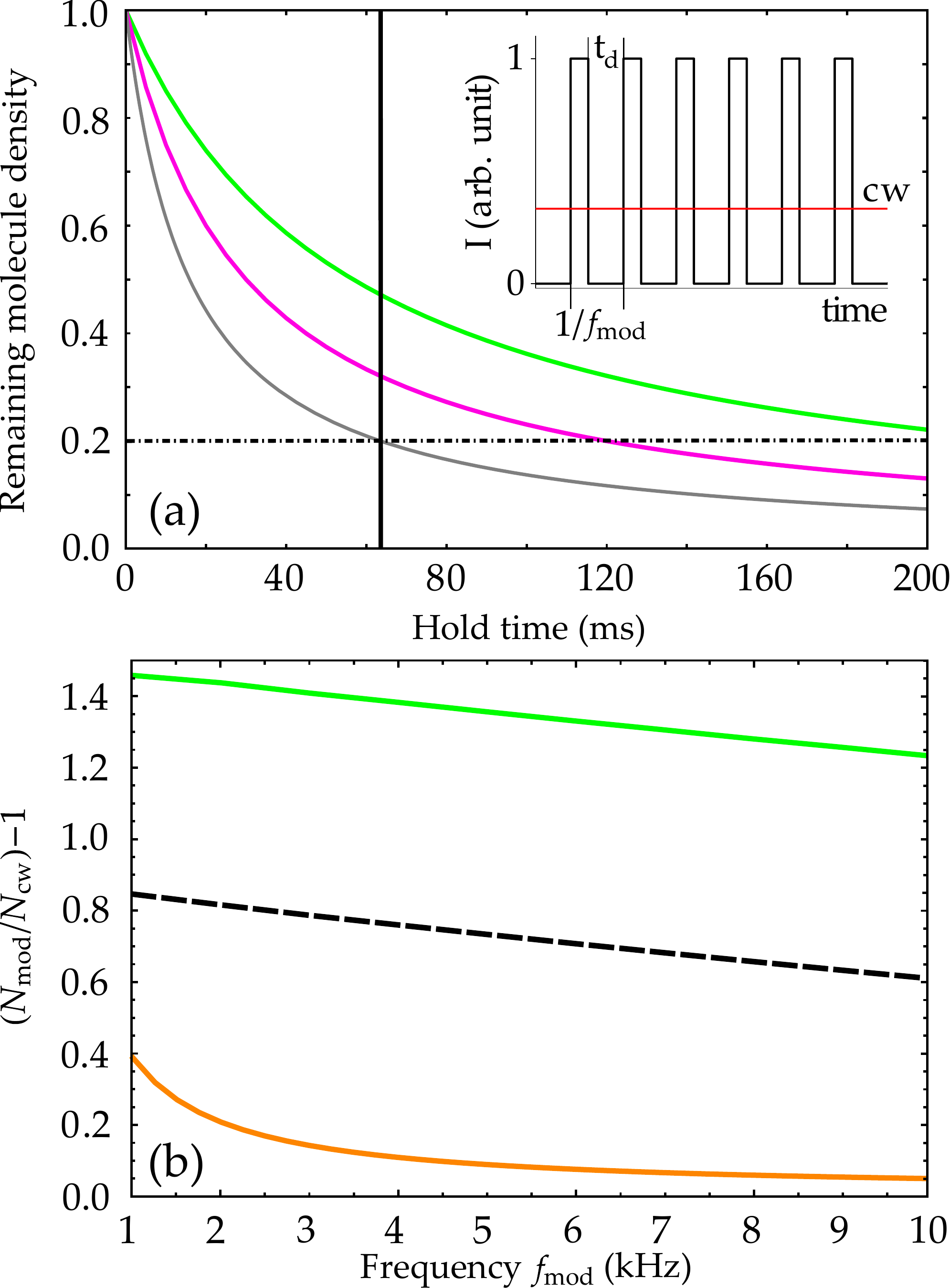}
\caption{ (a) Simulated \NaK\ two-body decay dynamics in a cw ODT (gray) in comparison to a chopped ODT with $f_\mathrm{mod}= 5\,$kHz and duty cycles $\eta$ of 0.5 (purple) and 0.25 (green). The horizontal line is a guide to the eye and marks the time (vertical line) at which 20\,\% of the molecules remain. The inset shows a sketch of the ODT sequence for the cw and chopped case. (b) Comparison of the predicted relative number gain between the cases of the different molecular species \NaK\ (solid green), \Na\Rb\ (dashed black) and \Rb\Cs\ (solid orange) dependent on the chopping frequency of the ODT. For \NaK\ and \Rb\Cs\ a density decrease to 20\,\% of the initial density is used, for \Na\Rb\ 30\,\%. The values and parameters are chosen according to experimentally realized data to provide direct comparison. In case of \Rb\Cs\ the parameters are chosen in conjunction with comparability to \Na\Rb\ and \NaK.}
\label{Modulation}
\end{figure}

In a first step, we apply the model to calculate the decay dynamics of molecules. Fig.\,\ref{Modulation}(a) shows loss curves for  \Na\K\ molecules. We chop the ODT intensity with a frequency $f_{\mathrm{mod}}=5\,$kHz and two different duty cycles $\eta$ of 0.25 and 0.5 resulting in $t_\mathrm{d}=(1-\eta)\times 1/f_\mathrm{mod}$ in which the ODT intensity is zero and the photoinduced loss term $k_l$ thus vanishes. For the simulation, we assume a complex lifetime of $6\,\mu$s and an initial molecular density of  $n_0=2\cdot10^{12}\,\text{cm}^{-3}$ at $t=0$, $k_2=4.49\times10^{-10}\,\mathrm{cm}^3\mathrm{s}^{-1}$ and $k_l=3.9\times 10^2\, \mathrm{s}^{-1}/\mathrm{W cm}^{-2}$\,\cite{PhysRevLett.123.123402}. As expected, we observe the molecular decay dynamics to significantly slow down with decreasing $\eta$. The effect is expected to become the stronger the slower the trap modulation frequency is. This is illustrated in Fig. \ref{Modulation}(b) showing the relative number gain $(N_\mathrm{mod}/N_{\mathrm{cw}})-1$ using a chopped ODT in comparison to a cw ODT. The green curve shows the result for \NaK\, at a hold time  (Fig. \ref{Modulation}(a) (vertical black line)). We also perform these simulations of the number gain for the \Na\Rb\ molecule and for a thorough comparison also for the \Rb\Cs\ molecule with realistic experimental parameters (black dashed and orange line). 
As can be seen from Fig. \ref{Modulation}(b), the relative number gain for \Na\Rb\ molecules and \Na\K\ is expected to reach values larger than 0.8 and 1 respectively, depending on the chosen parameters.  This is by far larger than the expected and observed relative number gain for the \Rb\Cs\ case (orange line in Fig. \ref{Modulation}(b)) \cite{PhysRevLett.124.163402} due to the much shorter estimated complex lifetimes for \Na\Rb\ and \Na\K\ .


To test the theoretical expectations, we perform lifetime measurements of both \Na\Rb\, and  \Na\K\, molecules in chopped ODTs. The experiments are performed in two different experimental apparatus, for \Na\Rb\, in Hong Kong and \Na\K\, in Hanover.

Both spin polarized ensembles of \Na\Rb\, and \Na\K\, molecules are prepared using Feshbach molecule association \cite{NaRbFeshbach,Voges2019Fesh} and transfer of the Feshbach molecules to a specific hyperfine state of the  $v=0, J=0$ rovibrational level of the $X^1\Sigma^+$ ground state using stimulated Raman adiabatic passage \cite{NaRb1,PhysRevLett.125.083401}. The \Na\Rb\, experiments start from an  ultracold molecular ensemble prepared in the lowest hyperfine state of the rovibrational ground-state manifold $\ket{J = 0, M_J = 0, m_{i,\text{Na}} =3/2, m_{i,\text{Rb}} =3/2}$~\cite{NaRb1}, with a typical temperature of $500\,$nK and a peak density of $5\times10^{11}\,\mathrm{cm}^{-3}$, where $J$ is the total molecular electronic spin and $M_J$ its projection, and $m_{i,\textrm{a}}$ the respective atomic nuclear spin projections. The \NaK\, experiments are performed with 
molecules in $\ket{J=0, M_J=0,m_{i,\text{Na}}=-3/2,m_{i,\text{K}}=-1/2}$ state, with a temperature of about 300\,nK and a peak density of $2\times10^{12}\,\mathrm{cm}^{-3}$.
Both ensembles are trapped in crossed ODTs operating at about $1064\,$nm and a mean trap frequency of about $2\pi\times100$\,Hz.


For both species, we implemented the chopped ODT with two methods. \\
(i) We used  double-passed AOM configurations to vary $f_{\mathrm{mod}}$ and $\eta$ arbitrarily, only limited by the rise and fall time of the AOM's resonance circuit. However, in this case the light extinction during $t_{\mathrm{d}}$ is not absolute. We measured  a leakage at the level of 100 ppb (10 ppb) for the \Na\Rb\, (\NaK) experiment. This corresponds to an intensity of $<5 \rm mW/cm^2$ for the typical ODT intensity in use. For both the \Rb\Cs\ \cite{PhysRevLett.124.163402} and the \40K\Rb\ \cite{liu_photo-excitation_2020} systems, such a low intensity causes nearly no loss of the complexes. We also ruled out light leakage from other laser sources as well as for radio frequency and microwave radiation.\\ 
(ii) We used an optical chopper wheel to totally block the light during $t_{\mathrm{d}}$. However, the drawback is somewhat restricted $f_{\mathrm{mod}}$ and $\eta$. Furthermore, the latency of the chopper also requires additional setup for switching between the cw ODT to the chopped one.

\begin{figure}[t]
\centering
\includegraphics[width=1.0 \linewidth]{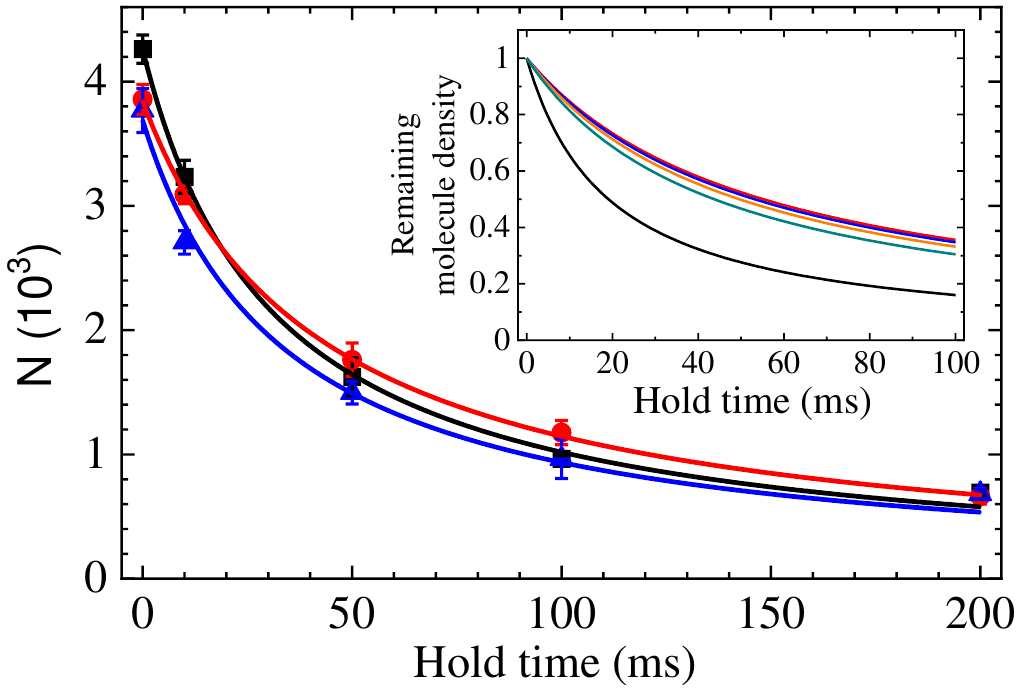}
\caption{Loss of \Na\Rb\, in the cw ODT (black squares) and chopped ODTs modulated at 1 kHz (red dots) and 2 kHz (blue triangles) with the optical chopper. The duty cycle is $\eta = 33\%$. Error bars show the standard error. Inset: Simulated \Na\Rb\, loss curves in the cw (bottom black curve) and chopped traps with $\eta = 33\%$. From top to bottom $f_{\rm mod}$ are red: 1 kHz, blue: 2 kHz, orange: 4 kHz, and cyan: 8 kHz for comparison.
}
\label{NaRbloss}
\end{figure}

As a first measurement, Fig.~\ref{NaRbloss} shows a direct comparison of the molecule number losses in the cw and chopped ODT for the case of \Na\Rb. In this experiment, the chopped ODT is modulated by the optical chopper at $\eta = 0.33$. The chosen $f_\textrm{mod}$ of 1 kHz and 2 kHz ensures that $t_\mathrm{d}$ is much longer than  the best known $\tau_\textrm{c,NaRb}=12.9\,\mu$s of $\rm Na_2Rb_2$~\cite{PhysRevA.100.032708,note1} which should allow the complex dissociation to happen. The initial sample conditions are nearly the same for the 3 measurements. However, we do not observe any sign of loss reduction in Fig.~\ref{NaRbloss}.  This is in contrast to the theoretical predictions (inset of Fig.~\ref{NaRbloss}) starting with similar initial sample conditions and $\tau_\textrm{c,NaRb}=12.9\,\mu$s, $k_2=3\times10^{-10}\,\mathrm{cm}^3\mathrm{s}^{-1}$ at a temperature $T=500\,$nK, and $k_l=4\times 10^2\, \mathrm{s}^{-1}/\mathrm{W cm}^{-2}$ (similar to RbCs~\cite{PhysRevLett.124.163402} and KRb~\cite{liu_photo-excitation_2020}) which clearly show an increasing suppression of two-body decay with decreasing modulation frequency.

A caveat for this direct comparison is the additional heating and also possibly one-body loss induced by the modulation. These obstacles have been ignored in the simulation. However, in this temperature range, $k_2$ for \Na\Rb\ molecules decreases with increasing collisional energy~\cite{ye18}. This should further enhance the loss reduction in the modulated trap which is contrary to our observation.

\begin{figure}[t]
\centering
\includegraphics[width=0.9 \linewidth]{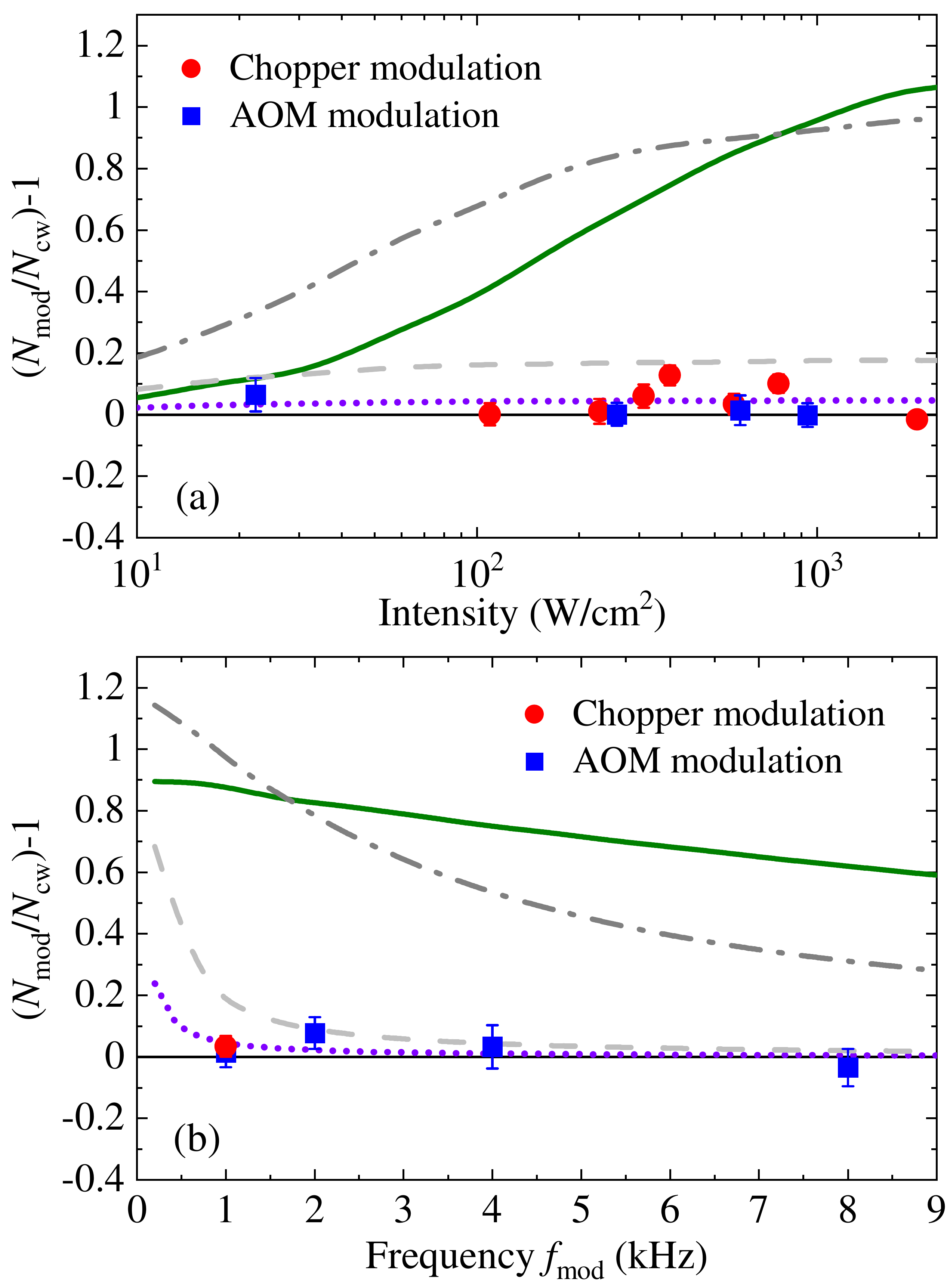}
 \caption{Probing the $\rm Na_2Rb_2$ complex in the chopped ODT with a killing beam. (a) Relative number gain after a holding time of 80 ms versus the killing intensity. The trap is modulated at $f_{\rm mod} = 1$~kHz and $\eta = 0.33$ with the two different methods. The killing beam wavelength is 1064.4 nm, same as for the ODT light. 
The green solid, grey dash-dotted, light gray dashed and purple dotted lines show the expected relative number gain as a function of cw killing intensity for different complex lifetimes $\tau_\textrm{c,NaRb}$ of 12.9 $\mu$s, 65.7 $\mu$s, 1 ms and 4 ms, respectively. 
 (b) Relative number gain as a function of  $f_{\rm mod}$ and thus $t_{\rm d}$ with a fixed killing intensity of 0.6 $\rm kW/cm^2$. The curves show the expected relative number gains for the several $\tau_\textrm{c,NaRb}$ in (a). The theoretical curvers are calculated with $k_l=4\times 10^2\, \mathrm{s}^{-1}/\mathrm{W cm}^{-2}$ (similar to RbCs~\cite{PhysRevLett.124.163402} and KRb~\cite{liu_photo-excitation_2020}).
 Error bars indicate $1\sigma$ standard errors.} \label{NaRbmod}
\end{figure}


\begin{figure}[t]
\centering
\includegraphics[width=0.87 \linewidth]{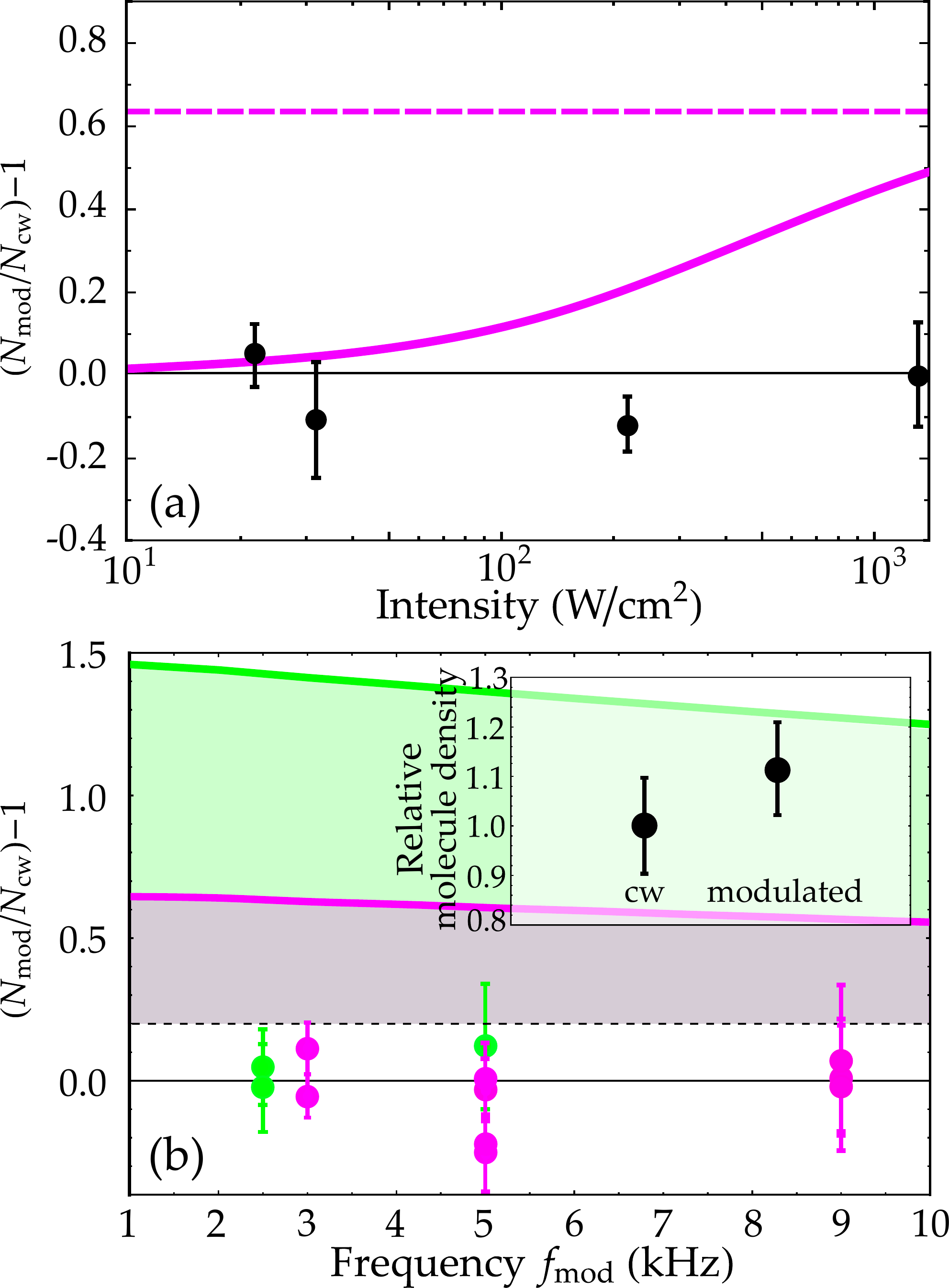}
 \caption{Probing the $\rm Na_2K_2$ complex in the chopped ODT with a killing beam. (a) Relative number gain at 20\,\% remaining density versus the killing beam intensity. The killing beam wavelength is 1064.5\,nm, the same as the ODT light. The trap is modulated via the chopper wheel at a frequency of $f_\mathrm{mod}=5\text{kHz}$ and a duty cycle $\eta=0.5$. The purple solid line shows the expected relative number gain for a complex lifetime of $6\,\mu\text{s}$. The dashed line represents the maximal achievable gain in the limit of high intensities. (b) Relative number gain as a function of modulation frequency for two different duty cycles $\eta=0.5$ (purple) or $\eta=0.25$ (green points). The purple(green) curve shows the expected gain at a duty cycle of $\eta=0.5$ ($\eta=0.25$). The value of all data points is expected to be above the horizontal dashed line and below the corresponding theoretical gain which should be resolvable for our experiments (green and purple shaded area).  Inset:  Selected experimental run for which the relative number gain is highest. The measurement has been done at $f_\text{mod}=5\,\text{kHz}$, $\eta=0.25$ and roughly $20\,\%$ of the initial molecule density left in the cw trap. Note, that measurements using the AOM technique are not marked  on purpose for clarity. Error bars show $1\sigma$ standard errors.}\label{NaKmod}
\end{figure}

To allow for a direct comparison and avoid the heating issues, for the following measurements, we mimic the cw ODT in a chopped ODT with an overlapped cw killing beam and make use of a large beam waist to minimally perturb the ODT potential. We now perform a systematic set of measurements for both \Na\Rb\, and \Na\K\, at a fixed holding time, measuring the remaining number of \Na\Rb(\NaK) molecules with the killing beam ($N_{\rm cw}$) and without it ($N_{\rm mod}$) and calculate the relative number gain $N_{\rm mod}/N_{\rm cw}-1$. We choose the holding time at 80\,ms(60\,ms) for \Na\Rb(\NaK) so that the relative number gain is nearly maximized in the simulation (compare Fig.~\ref{Modulation}(b) and Fig.~\ref{NaRbloss}).  In a first set of measurements, we keep $f_{\mathrm{mod}}$ fixed at 1\,kHz(5\,kHz) and $\eta$ at 0.33(0.5) for \Na\Rb(\NaK). At each intensity, we took at least 30 shots in alternatingly with and without the killing beam and calculated the relative number gain with the averaged numbers for each case\,\cite{PhysRevLett.124.163402}. Fig.  \ref{NaRbmod}(a) and Fig. \ref{NaKmod}(a) summarize the measurements for \Na\Rb\ and \NaK, respectively.  For \Na\Rb\ the relative number gain is observed to be consistent with zero independent of the chopping technique (chopper wheel (red dots)/AOM (blue squares)).  The same is true for \NaK, using the chopper wheel technique. This is in stark contrast to theoretical expectations.

To vary the effective dark time, we also measure the dependence of the relative number gain on $f_{\rm mod}$. This data is shown in Fig.~\ref{NaRbmod} (b) and Fig.~\ref{NaKmod}(b) for \Na\Rb\, and \NaK\, respectively. However, for both \Na\Rb\, and \NaK\, we are unable to observe any sign of frequency dependence. Instead, the relative number gain is constant over the whole frequency range and again consistent with zero. For the \NaK\, experiment, this even remains true when varying the  duty cycle from  $\eta=0.5$ (purple points) to $\eta=0.25$ (green points) --  again contrary to our expectations indicated by the purple(green) line for $\eta=0.5$($0.25$).

To exclude an unexpected weak coupling of the 1064\,nm killing beam, we also performed experiments on the \NaK\ molecule with different wavelength of 950\,nm and 816\,nm. An effect from these measurements remains similarly elusive as for all the other results.

The origin of this surprising result is at present unexplained. Assuming that photon scattering off the complexes from the ODT is indeed the dominating loss mechanism 
we can provide a lower bound for the complex lifetime: Taking the largest measured density gain from the \NaK\ data (see inset of Fig.~\ref{NaKmod}(b)), we extract a lower bound of $\tau_\textrm{c,NaK} = 0.35\,$ms. For \Na\Rb, this is $1\,$ms (light gray dashed curve in Fig.~\ref{NaRbmod}(b)). If we on the other hand consider all measurements to be equal, this would force $\tau_\textrm{c,NaK} > $1\,ms and $\tau_\textrm{c,NaRb} > $4\,ms.


In conclusion, we have probed photoinduced two-body loss of chemically stable bosonic \Na\Rb\ and  \NaK\ molecules in chopped ODTs. Given their small predicted complex lifetimes ~\cite{PhysRevA.100.032708,PhysRevLett.123.123402}, for both species we would have expected a strong suppression of the inelastic two-body loss far below the universal limit for easily accessible chopping frequencies and dark times. Instead, we observed near-universal decay of molecules independent of the chopping frequency and the dark time for both species. This is particularly surprising, since experiments have been done independently from one another in two different experimental apparatus in Hong Kong and Hanover. Furthermore, in parallel to our work similar results have been found in experiments with fermionic $^{23}$Na$^{40}$K by the Munich group \cite{Munich}.
Our results can in principle be interpreted in two ways: Either there is still some unknown about the complex lifetime or an unidentified loss mechanism dominates the molecular decay for the case of these relatively light-weighted molecular species. In any case, 
it will be crucial to find the origin of this surprising result. This is particularly important since universal two-body decay 
is believed to be understood for bialkali molecules following the proposals by Mayle \textit{et al.} \cite{PhysRevA.87.012709} and Christianen \textit{et al.} \cite{PhysRevA.100.032708,PhysRevLett.123.123402}, the  probing of photoinduced  loss with \Rb\Cs\ \cite{PhysRevLett.124.163402} and the measurement of  the transient intermediate complex lifetime for chemically reactive \40K\Rb\ \cite{liu_photo-excitation_2020}.

\textbf{Acknowledgements--}
The Hanover group thanks Leon Karpa for stimulating discussions and Jule Heier for laboratory assistance. The group gratefully acknowledges financial support from the Deutsche Forschungsgemeinschaft (DFG) through  the
collaborative research centre SFB 1227 DQ-mat,  the Research Unit 2247 project E05 and Germany’s Excellence Strategy – EXC-2123/1 QuantumFrontiers.  P.G. thanks the DFG for financial support through RTG 1991. The Hong Kong team thanks Hua Guo and Daiqian Xie for the new complex lifetime calculation, and Guanghua Chen and Mucan Jin for laboratory assistance. The team was supported by the Hong Kong RGC General Research Fund (grants 14301119, 14301818 and 14301815) and the Collaborative Research Fund C6026-16W. 

P.G./K.K.V. and J.L./J.H. contributed equally to this work. 


\bibliography{Strobo}

\begin{thebibliography}{30}
\bibitem [{not()}]{note1}%
  \BibitemOpen
  \href@noop {} {}\bibinfo {note} {By courtesy of Hua Guo and Daiqian Xie, a
  new $\rm Na_2Rb_2$ complex lifetime of $\tau_\textrm{c,NaRb} = 65.7~\mu$s is
  obtained from a latest quantum chemistry calculation.}\BibitemShut {Stop}%
\end{thebibliography}

\end{document}